\documentclass[a4paper,12pt]{article}
\usepackage{amssymb}
\usepackage{amsmath}
\usepackage{amsthm}
\usepackage{latexsym}
\begin{document}
\thispagestyle{empty}

\title{\textbf{Perturbative Computation of the Gluonic 
Effective Action via Polyakov's World-Line Path Integral}}
\author{\textbf{S. D. Avramis}
\thanks{e-mail: savramis@cc.uoa.gr},~~\textbf{~A. I. Karanikas}
\thanks{e-mail: akaranik@cc.uoa.gr}~~\textbf{and~C. N. Ktorides}
\thanks{e-mail: cktorid@cc.uoa.gr}}
\date{}
\maketitle
\begin{center}
\textit{University of Athens, Physics Department\\
Nuclear \& Particle Physics Section\\
Panepistimioupolis\\ GR 157--71, Athens, Greece}
\end{center}
\vspace{1cm}
\numberwithin{equation}{section}
\begin{abstract}
The Polyakov world-line path integral describing the propagation of gluon field quanta is 
constructed by employing the background gauge fixing method and is subsequently applied
to analytically compute the divergent terms of the one (gluonic) loop effective action  
to fourth order in perturbation theory. The merits of the proposed approach is that, 
to a given order, it reduces to performing two integrations, one over a set of Grassmann 
and one over a set of Feynman-type  parameters through which one manages to accommodate all 
Feynman diagrams entering the computation at once.
\end{abstract}
\newpage
\section{Introduction}
\vspace{.2cm}

Improved methods, in comparison to the Feynman diagrammatic ones, for expediting perturbative 
calculations in QCD have emerged, within the last decade or so, through the adoption of 
first-quantization-based approaches. The latter involve either string or world-line 
agents through which one describes the field theoretical system. The original 
efforts in this direction were string inspired and were based on realizations, made in 
the late 1980's [1-5], regarding the relation between string and non-abelian gauge field 
theories in the limit of an infinite string tension. Following their own involvement 
in such studies, Bern and Kosower [6] established a set of rules expediting efficient, 
one loop, computations in non-abelian gauge theories. Through them one could encompass 
contributions of a host of Feynman diagrams at once. Further extensions of the 
string-inspired approach were subsequently carried out in Refs. [7-12].

World-line based methodologies aiming at the same goals soon followed through the work 
of Strassler [13] who proposed suitably defined (one-dimensional) path-integrals for 
the various quantum field systems he considered. Extensive use was made of supersymmetric 
one-dimensional particle coordinates [14-16] `living' on the paths. The end result was the 
accomplishment of the full reproduction of the Bern-Kosower rules. Strassler's approach 
was further pursued in Refs. [17-19], where computations pertaining to multi-loop 
configurations in QED as well as effective actions involving constant, external 
(chromo)electric and (chromo)magnetic fields were undertaken.

Now, the world-line casting of relativistic quantum systems is an old story, which 
goes back to Fock [20], Feynman [21] and Schwinger [22]. Notable, relevant 
contributions followed by several authors [23-26], the latter of which sparked our 
original interest in the subject [27,28]. What particularly attracted our attention was the 
geometrical setting underlying the construction of Polyakov's (world-line) path integral. 
Within this context we pursued the problem of tracing the field theory origins of 
Polyakov's spin factor, introduced by him to properly account for the propagation 
of a free, spin-1/2 particle-like entity on a closed (Euclidean) space-time contour. 
In Ref. [28] we established, via a well-defined procedure, the emergence of the spin factor  
through the recasting of the, spin-1/2, matter field sector of a gauge field 
theory from a functional to a (world-line) path-integral, entering 
as an appropriate weight to account for spin. At the same time, the field theoretical 
interaction  term\footnote{The interaction term is displayed generically, independent as to 
whether the gauge theory is abelian or non-abelian.}  $\bar{\psi}\gamma_\mu  \psi A^\mu$  
is replaced by a `factorized' Wilson line (or loop) which accounts for the effect of the gauge
field on the world-line paths, equivalently describes its interaction with the matter 
particles. Our first applications turned in the direction of considering situations when 
it is justified to set the spin factor to unity -an occurrence which facilitates a 
factorization of the infra-red sector of the gauge field theory, in its perturbative 
version [29]. 

More recently, we have tested the possible merits stemming from the aformentioned 
disentanglement between spin-factor and Wilson line(loop), inherent in the Polyakov 
(world-line) path integral for spin-1/2 particle-like entities, as far as the task 
of facilitating effective, perturbative 
computations in QCD is concerned [30,31]. In the first of these papers the emphasis was 
placed on extending the world-line methodology to {\it open} fermionic lines. At the 
same time we established a procedure by which Strassler's path integral 
expression for spin-1/2 matter particles, entering an (interacting) gauge field theory and
which contains the term $\sigma\cdot F$ in the action,  can be recast into the Polyakov 
form which carries, in its place, the spin-factor. Our subsequent manipulations 
were expedited by the presence of the spin factor and produced, as a bonus, the 
following physical picture: Non-trivial spin-factor contributions come precisely at those
points where a gauge field quantum is emitted or absorbed by the fermionic world-line path.
Moreover, each such occurence signifies the presence of a derivative discontinuity on the path
as a four-momentum $k_\mu$ is locally injected. This is, indeed, a nice intuitive picture 
as it connects the mathematical fact pertaining to the dominance of non-differentiable
paths on the one hand, with the physical occurence of emission and absorption of quanta on 
the other. In the second 
paper we focused our attention on the more pragmatic goal of developing
algorithms, always for spin-1/2 particle (open) world-lines, which lead to efficient 
perturbative computations in QCD pertaining to Green's functions and amplitudes. Two 
different alternatives were arrived at according to
whether the execution of the particle path-integral preceeds or follows the considerations
involving the Wilson line (loop): (a) The Feynman diagrammatic logic is directly visible
and comprehesively dealt with, 
(b) a novel organization of the perturbative expansion is achieved which retains the space-time 
description all the way. Strategy (a) leads to a neat organization of the resulting 
perturbative expression which reduces the computation to two straightforward steps. 
The first pertains, basically, to the spin-factor and amounts to a 
simple integration over a set of Grassmann variables, as many in number 
as the perturbative order considered. The second integration 
is over a set of proper time parameters inherited from the expansion of the Wilson exponential.
Possible practical merits of alternative (b) constitute an open issue. 

Encouraged by the fact that the Polyakov path-integral for spin-1/2 particle entities
leads to computational procedures which,
both logistically and intuitively, seem to present advantages of their own, 
we undertake, in the present paper, the task of extending the application 
of the relevant methodology to the gluonic sector of QCD. Such an effort entails, among other 
things, the determination of the spin factor pertaining to the propagation of a spin-1 particle 
like entity. This issue will be confronted in Section 2 where we consider the pure gauge field
sector of a Yang-Mills system and utilize techniques associated with the background gauge
field fixing procedure. 
Focusing on effective action terms at the one gluon loop level we proceed, in Section 3, 
to produce a master expression 
furnishing the $Mth$ perturbative order contribution. The overall structure of these terms 
corresponds to a gluonic world-line 
loop on which `vertex operators', in the form of plane waves are attached. As in our 
previous work [31], pertaining to 
open fermionic world-lines, the overall calculation reduces to an integration over a 
Grassmann set of parameters followed by 
one over a set of Feynman-type parameters. The number of variables for each of the two 
sets is fixed by the perturbative order, 
while the execution of the integrals themselves is a fairly straightforward matter, 
as witnessed by the applications worked out in 
Section 4. There, the master formulas are used in order to compute the divergent parts of 
the 2nd, 3rd and 4th order one 
gluon loop contributions to the effective action. From the specific manipulations it 
becomes totally apparent how the 
world-line configuration accommodates, to a given order, the totality of the contributing 
Feynman diagrams via straightforward permutations of the parameters. As already 
mentioned, our analytic computations 
in this section are restricted to the divergent contributions (furnishing renormalization 
factors) which can be readily identified  
and isolated from the full expression to each given order. In this connection, 
it can be easily surmised from our master 
expressions that no divergent terms make their appearance above the 4th order, 
an occurrence that complies with the 
renormalizability of the theory. As far as the finite contributions are concerned, 
what we can say at this stage is that we are in the process of finalizing tests of relevant 
numerical procedures that have been devised for coping directly with their computation. 
We intend to report on this matter in the near future [32]. Finally, 
in Section 5 we summarize our findings 
and formulate our conclusions, while in an Appendix we trace the main steps 
involved in bringing the spin-factor to its final, ready to apply form.

\section{Polyakov World-Line Path Integral for the Gluon Sector of QCD}
\vspace{.2cm}
The successful trancription of the fermionic sector of a gauge field theory into its Polyakov
path integral\footnote{For convenience we drop the characterization `worldline' from hereon, 
even though we recognize the fact that we are using a term that has been established for
the characterization of the same author's path integral pertaining to string quantization.} 
form utilizes the fact that the corresponding functional integral is of a gaussian 
type [28]. For the gluon sector, of course, such is not the case. Following Refs. [13,19] we 
proceed by employing the background gauge fixing procedure according to which the gauge field
$A_\mu$ splits into a dynamical component, to be denoted by $\alpha_\mu$, and a
background field $B_\mu$. Given that we shall restrict, in the present paper, ourselves to the
computation of effective action terms, the background field will be considered as classical. 
Let us finally mention that we shall keep our formalism Euclidean throughout our analysis. 
Transcription of our final results to Minkowski space-time will be made in the end. In this
respect, characterizations such as `Lorentz generators', `Lorentz trace',
etc. will be employed by abuse of language.

The quadratic part of the (pure) gauge field action reads
\begin{equation}
S_2  = \frac{1}{2}\alpha _\mu ^a \left[ { - (D^2 )^{ab} \delta _{\mu \nu }  + (D_\nu  D_\mu  )
^{ab}  - \frac{1}{\xi }(D_\mu  D_\nu  )^{ab}  - igF_{\mu \nu }^{ab} } \right]\alpha _\nu ^b  
+ \bar c^a \left[ {(D^2 )^{ab} } \right]c^b, 
\end{equation}
where $D_\mu ^{ab}= D_\mu ^{ab} (B) = \partial _\mu  \delta ^{ab}+ gf^{abc} B_\mu$ is the 
covariant field derivative in the adjoint representation, while $c,\,\bar{c}$
are the ghost fields.  
Obviously, $F_{\mu\nu}$ entering Eq. (2.1) is the Maxwell tensor for the background 
gauge field, i.e
\begin{equation}
F_{\mu \nu }^{ab}  = F_{\mu \nu }^{ab} (B) =  - if^{abc} F_{\mu \nu }^c (B) =  - if^{abc} 
(\partial _\mu  B_\nu ^c  - \partial _\nu  B_\mu ^c  - gf^{cde} B_\mu ^d B_\nu ^e ).
\end{equation}

Adopting the Feynman gauge $(\xi=1)$ and introducing the Lorentz generators under which
four-vectors transform, namely
\begin{equation}
\left( {J_{\rho \sigma } } \right)_{\mu \nu }  = i(\delta _{\rho \mu } 
\delta _{\sigma \nu }  - \delta _{\rho \nu } \delta _{\sigma \mu } ),
\end{equation}
we rewrite Eq. (2.1) as follows
                                                           
\begin{equation}
S_2  = \frac{1}{2}\alpha _\mu ^a \left[ { - (D^2 )^{ab} \delta _{\mu \nu }  
- g(J \cdot F)_{\mu \nu }^{ab} } \right]\alpha _\nu ^b  + \bar c^a \left[ {(D^2 )^{ab} }
 \right]c^b. 
\end{equation}

In the one loop approximation, to which we shall restrict our considerations in this work,
the effective action, as a functional of the background field, is given by
\begin{equation}
\Gamma _1 \left[ B \right] = \frac{1}{2}Tr\ln \left( { - D^2  - gJ \cdot F} \right) - 
Tr\ln ( - D^2 )= \Gamma _{1,gluons} \left[ B \right] + \Gamma _{ghosts} \left[ B \right].
\end{equation}                              
In what follows it suffices to work with $\Gamma _{1,gluons} \left[ B \right]$ as 
$\Gamma _{ghosts} \left[ B \right]$ is simply given by the first of the two terms 
entering the gluon contribution to the effective action, multiplied by (-2). 

Employing Schwinger's parametrization formula [22] we write (trace on `Lorentz' and
color indices)
\begin{equation}
\Gamma _{1,gluons} \left[ B \right] =  - \frac{1}{2}\int\limits_0^\infty  
{\frac{{dT}}{T}} \int {d^D } xTrK(x,x;T),
\end{equation}
where                                     
\begin{equation}
K(y,x;T)_{\mu \nu }^{ab}  \equiv \left\langle y \right|e^{ - T( - D^2  - gJ \cdot F)} 
\left| x \right\rangle _{\mu \nu }^{ab} 
\end{equation}
corresponds to the (dynamical) gauge field propagator kernel in the background field.

The world-line path integral for $K(y,x;T)_{\mu \nu }^{ab}$ results through standard 
procedures (see, e.g., Ref. [30]) and reads                                          
\begin{equation}
K(y,x;T)_{\mu \nu }^{ab}  = \int\limits_{x(0) = x,x(T) = y} {Dx(t)\exp \left[ { - 
\frac{1}{4}\int\limits_0^T {dt\dot x^2 (t)} } \right]} P\exp \left[ {ig\int\limits_0^T 
{dt\dot x \cdot B + g\int\limits_0^T {dtJ \cdot F} } } \right]_{\mu \nu }^{ab}. 
\end{equation}
       
As already established in Refs. [30,31] the Polyakov path integral results once we apply the
`area derivative' operator [33,34] given by
\begin{equation}
\frac{\delta }{{\delta s_{\mu \nu } (t)}} \equiv \mathop {\lim }\limits_{\varepsilon  \to 0} 
\int\limits_{ - \varepsilon }^\varepsilon  {ds\,s\frac{{\delta ^2 }}{{\delta x_\mu  (t + 
\frac{s}{2})\delta x_\nu  (t - \frac{s}{2})}}} 
\end{equation}
and use, at the same time, the identities                                                 
\begin{equation}
\frac{\delta }{{\delta s_{\mu \nu } (t)}}P\exp (ig\int\limits_0^T {dt\dot x \cdot B)^{ab} }  
= {\rm P}\exp (ig\int\limits_t^T {dt\dot x \cdot B)_{\mu \rho }^{aa_1 } } 
\left({- igF(x(t))} \right)_{\rho \sigma }^{a_1 a_2 } {\rm P}\exp (ig\int\limits_0^t 
{dt\dot x \cdot B)_{\sigma \nu }^{a_2 b} } 
\end{equation}
and          	
\begin{equation}
\int\limits_0^T dt\frac{\delta }{{\delta s_{\mu \nu } (t)}}\exp ( - \frac{1}{4}
\int\limits_0^T {dt\dot x^2 ) 
= \frac{1}{2}} \int\limits_0^T dt\,
\omega _{\mu \nu } \left[ \dot{x}(t) \right]\exp ( - \frac{1}{4}\int\limits_0^T {dt\dot x^2 } ),
\end{equation}
where $\omega_{\mu\nu}$ expresses the rotation of the vector tangent to the trajectory [26]
and for paths described by differentiable functions assumes the form                           
\begin{equation}
\omega _{\mu \nu }[\dot{x}]= \frac{T}{2}(\ddot x_\mu\dot x_\nu -\dot x_\mu \ddot x_\nu).
\end{equation}                                                               
A more careful discussion pertaining to the spin factor is conducted in the appendix.

Once performing a partial integration, Eq. (2.8) assumes its Polyakov path-integral form
which reads              
\begin{equation}
K(y,x;T)_{\mu \nu }^{ab}  = \int\limits_{x(0) = x,x(T) = y} {Dx(t)\exp( - \frac{1}{4}
\int\limits_0^T {dt\dot x^2 }}) {\rm P}\exp(\frac{i}{2}\int\limits_0^T {dtJ \cdot 
\omega)_{\mu \nu }}{\rm P}\exp(ig\int\limits_0^T {dt\dot x \cdot B)^{ab} }. 
\end{equation}
In turn, the corresponding expression for the effective action, including the contribution
from the ghost term, becomes      
\begin{equation}                   
\Gamma _1 \left[ B \right] =  - \frac{1}{2}\int\limits_0^\infty  {\frac{{dT}}{T}} 
\int\limits_{x(0) = x(T)} {Dx(t)\exp ( - \frac{1}{4}} \int\limits_0^T {dt\dot x^2 } )
\left\{ {Tr_L \Phi ^{\left[ 1 \right]} \left[ {\dot x} \right] - 2} \right\}Tr_c
P\exp (ig\int\limits_0^T {dt\dot x \cdot B)}, 
\end{equation}
where the indices on the traces stand for `Lorentz' ($L$) and `color' ($c$) and where           
\begin{equation}
\Phi^{[1]}\left[\dot{x}\right]_{\mu\nu}\equiv {\rm P}\exp 
\left[ {\frac{i}{2}\int\limits_0^T {dtJ \cdot \omega [\dot{x}(t)]}}\right]_{\mu\nu } 
\end{equation}
is the spin-factor expression\footnote{Strictly speaking, Polyakov's spin-factor is given 
by the (Lorentz) trace of $\Phi^{[1]}\left[\dot{x}\right]_{\mu\nu}$.} for the spin-1 
particle-like entity (gluon) propagating in (Euclidean)space-time. It is not
difficult to see that the spin-factor has a restricted dependence on a path's profile.
As argued in Ref. [30] and further deliberated on in the appendix, contributions 
of the spin-factor to the path integral comes solely from points where a four-momentum is 
applied through an emission or absorbtion of a gauge field quantum. Roughly speaking, this
has to do with the fact that the expectation value 
$<\ddot{x}_\mu\dot{x}_\nu>-<\dot{x}_\mu\ddot{x}_\nu>$, as computed through the 
path integral, vanishes unless a four-momentum $k_\mu$ is imparted at the point $x$. 
                                             
For the sake of comparison we give the corresponding expression for the one fermionic loop
expression which contributes to the effective action [28]. It reads (color matrices in the 
fundamental representation) 
\begin{equation}
\Gamma _{1,fermions} \left[ B \right] = \frac{1}{2}\int\limits_0^\infty  
{\frac{{dT}}{T}} \int\limits_{x(0) = x(T)} {Dx(t)\exp ( - \frac{1}{4}} 
\int\limits_0^T {dt\dot x^2 } )Tr_L \Phi ^{\left[ {1/2} \right]} \left[ {\dot x} 
\right]Tr_c P\exp (ig\int\limits_0^T {dt\dot x \cdot B)} 
\end{equation}
with the spin factor now given by                          
\begin{equation}
\Phi ^{\left[ {1/2} \right]} \left[ {\dot x} \right]_{}  \equiv {\rm P}\exp 
\left[ {\frac{i}{2}\int\limits_0^T {dt\,S \cdot \omega [\dot{x}(t)]} } \right]
\end{equation}
and where the corresponding Lorentz generators belong to the spinor representation, i.e.
\begin{equation}                                                  
S_{\mu \nu }  = \frac{1}{2}\sigma _{\mu \nu }  = \frac{i}{4}\left[ {\gamma _\mu ,
\gamma _\nu  } \right].
\end{equation}
Generally put, the Polyakov path integral recasting of a relativistic, quantum field
theoretical system provides a unified basis for the description of the propagating
particle-like entity; simply one has to adjust the weight provided by the spin-factor 
to its particular form. Thus,
for a spin-zero particle the relevant weight factor is, simply, unity (note,
in this regard, that ghosts fall into this class irrespective of the anticommutation 
relations, cf. minus sign, they obey) while for spin-1/2 and spin-1 particle-like entities 
the corresponding weights are provided by Eqs. (2.17) and (2.15), respectively.

For completeness let us mention that the path integral expression 
for the gluonic Green's function, namely                                                           
\begin{equation}
iG(y,x)_{\mu \nu }^{ab}  = \int\limits_0^\infty  {dTK(y,x;T)_{\mu \nu }^{ab} }, 
\end{equation}
is determined once the substitution from Eq. (2.13) is made for the propagation kernel. 

\vspace{.5cm}
\section{The one Gluon Loop, M-point Effective Action}
\vspace{.2cm} 

In this Section we shall perform a number of manipulations through which
we shall arrive to ready-to-apply master expressions for the computation of one-loop
effective action terms. Let us commence our efforts by
substituting plane wave solutions of the (classical) Yang-Mills equations satisfied by
the background field, i.e. setting $B_\mu (x)={t_G^{a_n }\varepsilon _\mu ^n } e^{ip_n  
\cdot x}$ for each gauge field entering the M-th order term in the expansion of the
Wilson exponential in (2.14). We obtain 

\begin{eqnarray}
 &&\Gamma _1^{(M)} (p_1 ,...,p_M ) =  - \frac{1}{2}(ig)^M Tr_C (t_G^{a_M } ...t_G^{a_1 } )
\int\limits_0^\infty  {\frac{{dT}}{T}} \left[ {\prod\limits_{n = M}^1 {\int\limits_0^T 
{dt_n } } } \right]\theta (t_M ,...,t_1 )\nonumber\\ &&  
  \times \int\limits_{x(0) = x(T)} {Dx(t)\prod\limits_{n = M}^1 {\varepsilon ^n } }  
\cdot \dot x(t_n )\left\{ {Tr_L \Phi ^{\left[ 1 \right]} \left[ {\dot x} \right] - 2} 
\right\}\exp \left[ { - \frac{1}{4}\int\limits_0^T {dt\dot x^2  + 
i\sum\limits_{n = 1}^M {p_n  \cdot x(t_n )} } } \right]\nonumber\\&&  
  + permutations, 
 \end{eqnarray}
where $\theta (t_M ,...,t_1 ) =\prod\limits_{n = M - 1}^1 {\theta (t_{n + 1} - t_n })$ and
where the indication $permutations$ refers to all possible rearrangements of  the $t_n$
and the $t_G^{a_n}$ associated with the, $M$ in number, background gauge fields.

Our computational strategy for confronting the above quantity coincides
with the one employed in Ref. [31]. It relies on a move to recast the spin-factor expression
into an explicitly path-independent form. Once this is done the path integration can be
immediately performed given that the `action functional' is a simple gaussian 
(with a linear term). Subsequently, we shall deal with the spin-factor.

Following the procedure employed in the aformentioned reference we introduce the Grassmann 
variables $\bar{\xi}_n$ and $\xi_n$ through which the $\varepsilon ^n \cdot \dot x(t_n )$ 
factors in Eq. (3.1) are elevated into exponentials according to [13]
\begin{equation}
i\varepsilon ^n  \cdot \dot x(t_n ) = \int {d\xi _n } d\bar \xi _n \exp \left[ {i\xi _n 
\bar \xi _n \varepsilon ^n  \cdot \dot x(t_n )} \right].
\end{equation}
After substituting in (3.1) we obtain                                                     
\begin{eqnarray}
&&\Gamma _1^{(M)} (p_1 ,...,p_M ) =  - \frac{1}{2}g^M Tr_C (t_G^{a_M } ...t_G^{a_1 } )
\int\limits_0^\infty  {\frac{{dT}}{T}} \left[ {\prod\limits_{n = M}^1 {\int\limits_0^T 
{dt_n } } } \right]\theta (t_M ,...,t_1 )\left[ {\prod\limits_{n = M}^1 {\int {d\xi _n 
d\bar \xi _n } } } \right]\nonumber\\ && \times \int\limits_{x(0) = x(T)} {Dx(t)
\left\{ {Tr_L \Phi ^{\left[ 1 \right]} \left[ {\dot x} \right] - 2} \right\}} \exp 
\left[ { - \frac{1}{4}\int\limits_0^T {dt\dot x^2  + i\sum\limits_{n = 1}^M {\hat k(t_n ) 
\cdot x(t_n )}}}\right]\nonumber\\&& + permutations, 
\end{eqnarray}
having set               
\begin{equation}
\hat k_\mu  (t_n ) \equiv p_{n,\mu }  + \bar \xi _n \xi _n \varepsilon _\mu ^n 
\frac{\partial }{{\partial t_n }}.
\end{equation}

Recalling the original specification of the spin-factor, which utilizes the employment of the 
area derivative, let us rewrite Eq. (3.3) in the specific form which takes into account the 
fact that the gauge potentials entering the expansion of the Wilson exponential are plane 
waves. For the $Mth$ order term we write 
\begin{eqnarray}
Tr_L \Phi ^{\left[ 1 \right]} \left[ {\hat k} \right] &\equiv &\exp 
\left[ { - i\sum\limits_{n = 1}^M {\hat k(t_n ) \cdot x(t_n )} } \right]
\left[ {Tr_L {\rm P}\exp \left( { - i\int\limits_0^T {dtJ \cdot 
\frac{\delta }{{\delta s}}} } \right)} \right]\nonumber\\&&\times
\exp \left[ {i\sum\limits_{n = 1}^M {\hat k(t_n ) \cdot x(t_n )} } \right].
\end{eqnarray}
The above expression illustrates in an immediate, albeit formal, manner the path-independence 
of the spin-factor: One observes that 
the area derivative acting on the exponential will produce delta functions entering
the parametric integration, cf. Eq. (2.9), entering the definition of the area derivative 
operator. A well defined argument leading to this result is provided in the appendix.

Returning to the case in hand, we write   
\begin{eqnarray}
&&\Gamma _1^{(M)} (p_1 ,..,p_M ) =  - \frac{1}{2}g^M Tr_C (t_G^{a_M } ...t_G^{a_1 } )
\int\limits_0^\infty  {\frac{{dT}}{T}} \left[ {\prod\limits_{n = M}^1 {\int\limits_0^T 
{dt_n } } } \right]\theta (t_M ,...,t_1 )\left[ {\prod\limits_{n = M}^1 {\int {d\xi _n 
d\bar \xi _n } } } \right] \nonumber\\&& 
  \times \left\{ {Tr_L \Phi ^{\left[ 1 \right]} \left[ {\hat k} \right] - 2} \right\}
\int\limits_{x(0) = x(T)} {Dx(t)} \exp \left[ { - \frac{1}{4}\int\limits_0^T {dt\dot x^2  
+ i\sum\limits_{n = 1}^M {\hat k(t_n ) \cdot x(t_n )} } } \right]\nonumber\\&&
+ permutations.  
\end{eqnarray}
The first task we shall carry out is to perform the, basically Gaussian, path integral. 
Straightforward manipulations, partly displayed in the appendix, lead to the result
\begin{eqnarray}
&& \int\limits_{x(0) = x(T)} {Dx(t)\exp \left[ { - \frac{1}{4}\int\limits_0^T {dt
\dot x^2 (t) + i\sum\limits_{n = 1}^M {\hat k(t_n ) \cdot x(t_n )} } } \right]}  = 
(2\pi )^D \delta ^{(D)} (\sum\limits_{n = 1}^M {p_n } )\frac{1}{{(4\pi T)^{D/2} }} \nonumber\\ 
&& \times \exp \left[ {\sum\limits_{n < m} {p_n  \cdot p_m G(t_n ,t_m ) + 
\sum\limits_{n \ne m} {\bar \xi _n \xi _n \varepsilon ^n  \cdot p_m \partial _n 
G(t_n ,t_m )}}}\right.\nonumber\\&&
+ \left.\frac{1}{2}\sum\limits_{n \ne m} {\bar \xi _n \xi _n \bar \xi _m \xi _m 
\varepsilon ^n  \cdot \varepsilon ^m \partial _n \partial _m G(t_n ,t_m )} \right].  
 \end{eqnarray}
One notes contributions pertaining solely to the points of attachment of external gauge
field on the loop contour. In the above expression the following Green's function [13] has
been employed
\begin{equation}
G(t,t') = \left| {t - t'} \right|\left[ {1 - \frac{{\left| {t - t'} \right|}}{T}} \right].
\end{equation}
It corresponds\footnote{To be exact, the solution of Eq. (3.9) includes an arbitrary
constant which does not appear in Eq. (3.8) on the account that momentum conservation has 
been imposed.} to the motion of a one-dimensional particle moving on a closed 
contour, i.e.
\begin{equation}         
\frac{1}{2}\frac{{\partial ^2 }}{{\partial t^2 }}G(t,t') =  - \delta (t - t') + \frac{1}{T}
\end{equation}
and obeys the boundary conditions $G(0,t') = G(T,t')$ and $\dot G(0,t') = \dot G(T,t')$.

Introducing the dimensionless parameters $u_i$ according
to $t_i  = Tu_i ,i = 1,...,n$, the interim result for $\Gamma _1^{(M)} (p_1 ,...,p_M )$ reads      
\begin{eqnarray}
&& \Gamma _1^{(M)} (p_1 ,...,p_M ) =  - \frac{1}{2}g^M (2\pi )^D \delta ^{(D)}
 (\sum\limits_{n = 1}^M {p_n } )Tr_C (t_G^{a_M }  \cdot  \cdot  \cdot t_G^{a_1 } )
\frac{1}{{(4\pi )^{D/2} }}\int\limits_0^\infty  {dT} T^{M - D/2 - 1}  \nonumber\\&& 
  \times \left[ {\prod\limits_{n = M}^1 {\int\limits_0^1 {du_n } } } \right]
\theta (u_M ,...,u_1 )F^{(M)} (u_1 ,...,u_M ;T)\exp \left[ {T\sum\limits_{n < m}^{} {p_n  
\cdot p_m G(u_n ,u_m )} } \right] \nonumber\\&&\quad+ permutations, 
 \end{eqnarray}
where $G(u_n ,u_m ) = \left|{u_n- u_m }\right|\left[ {1 - \left| {u_n- u_m }\right|} \right]$ 
satisfies the additional properties
\begin{equation}
\partial _n G(u_n ,u_m ) \equiv \dot G(u_n ,u_m ) = sign(u_n  - u_m ) - 2(u_n  - u_m ) =
  - \dot G(u_m ,u_n )
\end{equation}                           
and
\begin{equation}
- \partial _n \partial _m G(u_n ,u_m ) = \partial _n^2 G(u_n ,u_m ) \equiv \ddot 
G(u_n ,u_m ) = 2\left[ {\delta (u_n  - u_m ) - 1} \right].
\end{equation}
Finally, in Eq. (3.10) we have set
\begin{eqnarray}
&&F^{(M)} (u_1 ,...,u_M ;T) = \left[{\prod\limits_{n = M}^1{\int{d\xi _n d\bar \xi _n }}}
 \right]\left( {Tr_L \Phi ^{\left[ 1 \right]}\left[\hat{k}\right]- 2}\right) \nonumber\\&& 
  \times \exp \left[ {\sum\limits_{n \ne m} {\bar \xi _n \xi _n \varepsilon ^n  
\cdot p_m \partial _n G(u_n ,u_m )}}\right.\nonumber\\&& 
+\left. \frac{1}{{2T}}{\sum\limits_{n \ne m} 
{\bar \xi _n \xi _n \bar \xi _m \xi _m \varepsilon ^n  \cdot \varepsilon ^m 
\partial _n \partial _m G(u_n ,u_m )} } \right]. 
\end{eqnarray}

The spin-factor can now be brought into a ready-to-apply form through a series
of manipulations that are outlined in the appendix. The following result is arrived at                    
\begin{eqnarray}
&&\Phi _{\mu \nu }^{\left[ 1 \right]}\left[\hat{k}\right]= {\rm P}\exp \left[ {\frac{i}{2}
\sum\limits_{n = 1}^M {J \cdot \phi (n)} } \right]_{\mu \nu } 
= \delta _{\mu \nu }  + \frac{i}{2}(J_{\rho \sigma } )_{\mu \nu } 
\sum\limits_{n = 1}^M{\phi_{\rho\sigma}}(n)\nonumber\\&& \quad +\left(\frac{i}{2}\right)^2 
(J_{\rho _2 \sigma _2 })_{\mu \lambda } (J_{\rho _1 \sigma _1 } )_{\lambda \nu } 
\sum\limits_{n_2  = 1}^M {\sum\limits_{n_1  = 1}^{n_2- 1}{\phi _{\rho _2 \sigma _2 }(n_2 )
\phi _{\rho _1 \sigma _1 } (n_1 ) + ...} }   
 \end{eqnarray}
with                   
\begin{equation}
\phi _{\mu \nu } (n) = 2\bar \xi _n \xi _n (\varepsilon _\mu ^n p_{n,\nu }  - 
\varepsilon _\nu ^n p_{n,\mu } ) + \frac{4}{T}\bar \xi _{n + 1} \xi _{n + 1} 
\bar \xi _n \xi _n (\varepsilon _\mu ^{n + 1} \varepsilon _\nu ^n  - 
\varepsilon _\nu ^{n + 1} \varepsilon _\mu ^n )\delta (u_{n + 1}  - u_n )
\end{equation}
and where we have designated that $\bar{\xi}_{M+1}=\xi_{M+1}=0$.

Two observations of practical interest can be made in connection with the above
expression for the spin-factor. First, it is clear that the number of terms in the
expansion of the exponential in Eq. (3.14) terminates at $M$ as the saturation point of
the Grassman variables is by then reached. Second, the delta function containing term in (3.14)
implies that for a given ordering there is a contribution from coinciding points,
$u_n$ and $u_{n+1}$ in this case. This occurrence signifies the presence of a `four-gluon
vertex' which is automatically included in a given perturbative calculation, along with
the (derivative-dependent) `three-gluon vertices' represented by the first term. One 
thereby concludes that the $Mth$ order perturbative contributions to the effective action
are classified, via the spin-factor, exclusively by the number of the points 
of gluon (single or pairwise) attachements on the closed world-line contour, in all possible
permutations. Accordingly, the computation of the M-point effective action term will collect 
all $Mth$ order, 1PI Feynman diagrams.

We mention in passing that fermionic loop contributions to the effective action easily follow
by referring to Eqs. (2.16)-(2.18). One simply has to make the substitution 
$\phi_{\rho\sigma}(J_{\rho\sigma})_{\mu\nu}\rightarrow S_{\mu\nu}\phi_{\mu\nu}$ in
Eqs. (3.14) and (3.15) and use the fundamental representation of the group.

Given the expressions we have arrived at, what remains to be carried out are the integrations 
over the Grassmann variables as well as the parametric integrations entering Eq. (3.10).  
Numerical methods have been developed for this purpose whose report is 
forthcoming [32]. For the rest of this paper we restrict ourselves to the 
computation of the divergent part of the effective action. In this regard, let us
observe, by looking at Eq. (3.13), that ultraviolet divergencies will occur only for 
terms of order $M=2,3,4$. Specifically, by focusing on the terms that have the minimum
number of $p_{n,\mu}$ factors one determines, through dimensional considerations, that they 
should carry the compensating factor $T^{2-M}$ for $M=2,3,4$. The latter, combines with
$T^{M-1-D/2}$ in (3.10) to produce divergent terms $\sim\Gamma\left(2-{D\over 2}\right)$.
Further inspection shows that no such terms arise for $M\geq 5$, a fact that directly complies
with the renormalizability of the theory. 

As our final remark in this Section, let us mention that our manipulations up to now have
treated the gluons as `real', i.e. we have assumed that                       
\begin{equation}
\varepsilon ^n  \cdot p_n  = 0\quad {\rm and}\quad p_n^2  = 0, \quad n = 1,2,\cdot\cdot\cdot.
\end{equation}
In the following Section, where our computations will run against infra-red problems, we 
shall protect our expressions from corresponding divergencies by going slightly off mass shell.
i.e. by setting $p_n^2  = \lambda ^2$, with $\lambda^2>\Lambda^2_{QCD}$. 

\vspace{.5cm}
\section{Computation of Divergent one-loop Effective Action Terms to Fourth Order}
\vspace{.2cm}
                                                                                                                   
In this Section we shall apply our comprehesive formulas given by Eqs. (3.10) and 
(3.13)-(3.15) towards the computation of the divergent contributions to the $M=2,3,4$
terms in the expansion of the effective action -in fact, the only terms which exhibit
ultra-violet divergencies. We leave the task of computing finite contributions, to the
same order, to a future paper where numerical methods will be applied.

\vspace{.5cm}
{\bf 4.1 The two gluon contribution (M=2) to the Effective Action}
\vspace{.3cm}

The present calculation pertains to the one gluon loop configuration with two
(truncated) gluon and ghost attachments.
Our master expression  accomodates all the contributing Feynman diagrams.
From Eqs. (3.14) and (3.15) we determine, for $M=2$,
\begin{equation}
Tr_L \Phi ^{\left[ 1 \right]}  = D - 8\bar \xi _1 \xi _1 \bar \xi _2 \xi _2 
\varepsilon ^1  \cdot \varepsilon ^2 p_1  \cdot p_2. 
\end{equation}
Upon substituting in (3.13) and performing the Grassmann integrations we obtain                                       
\begin{equation}
F^{(M = 2)} (u_1 ,u_2 ;T) =  - \frac{1}{T}(D - 2)\varepsilon ^1  \cdot 
\varepsilon ^2 \ddot G(u_1 ,u_2 ) - 8\varepsilon ^1  \cdot \varepsilon ^2 p_1  \cdot p_2. 
\end{equation}
One notes that the delta function entering the specification of $\ddot{G}(u_1,u_2)$ 
accomodates the contribution coming from the class of Feynman diagrams
wherein the two (truncated) external gluons attach themselves to the loop
through a four-point vertex. 
                
The above result when substituted in (3.13) gives, after an integration by parts which
results in the replacement $\ddot G(u_1 ,u_2 )\to - Tp_1\cdot p_2 \dot G^2(u_1 ,u_2 )$,
\begin{eqnarray}
&&\Gamma _1^{(M = 2)} (p_1 ,p_2 ) =  - \frac{1}{2}(2\pi )^4 \delta ^{(4)} (p_1  + p_2 )
Tr_C (t_G^{a_2 } t_G^{a_1 } )\frac{{g^2 }}{{(4\pi )^{D/2} }}\varepsilon ^1  \cdot 
\varepsilon ^2 p_1  \cdot p_2  \nonumber\\&& 
  \times \int\limits_0^\infty  {dTT^{1 - D/2} } \int\limits_0^1 {du_2 } \int\limits_0^{u_2 }
 {du_1 } \left[ {(D - 2)\dot G^2 (u_1 ,u_2 ) - 8} \right]\exp \left[ { - T\lambda ^2 
G(u_1 ,u_2 )} \right]\nonumber\\&&+ permutations,  
 \end{eqnarray}
where the infra-red cutoff $\lambda$ has been introduced by going off shell.
The integrations in the last equation can be easily performed and lead to the final result      
\begin{eqnarray}
\Gamma _1^{(M = 2)} (p_1 ,p_2 ) &=&  - \frac{1}{2}(2\pi )^4 \delta ^{(4)} (p_1 + p_2 )
N\delta ^{a_2 a_1 } \frac{{g^2 }}{{(4\pi )^2 }}\left( {4\pi \frac{{\mu ^2 }}{{\lambda ^2 }}} 
\right)^{2 - D/2} \varepsilon ^1  \cdot \varepsilon ^2 p_1  \cdot p_2  \nonumber\\&& 
  \times \Gamma \left(2 - \frac{D}{2}\right)\frac{{11 - 7(2 - D/2)}}{{3 - 2(2 - D/2)}}
B\left(\frac{D}{2} - 1,\frac{D}{2} - 1\right),  
 \end{eqnarray}
where the adjustment $g^2 \to g_D ^2 = g^2 \mu ^{4 - D}$ was made in order to restore
dimensional consistency. The term $permutations$ in Eq. (3.10) has been duly
taken care of by taking into account all the rearrangements of indices (1,2) and dividing 
by 2! in order to comply with boson non-distinguishability.

From Eq. (4.4) we verify, once returning to Minkowski space-time, the well known result (which
does not take into account the contribution from the fermionic loop) 
\begin{equation}
Z_g  = 1 - \frac{1}{2}\frac{{g^2 }}{{(4\pi )^2 }}N\frac{{11}}{3}\frac{1}{{2 - D/2}}.
\end{equation}
It is of interest to observe, by referring to Eqs. (2.13), (2.19) and as has been
explicitly demonstrated in Refs. [30,31], that the corresponding formulas resulting from 
Polyakov's path-integral for open lines have the same basic structure with the ones that 
have resulted from the present considerations pertaining to loops. It, then, becomes a straight
forward matter to surmise the validity of Ward's identity $Z_B ^{1/2} Z_g = 1$ which is known
to hold in the framework of the background gauge fixing method.

\vspace{.5cm}
{\bf 4.2 The three gluon contribution (M=3) to the Effective Action}
\vspace{.3cm}

We now turn our attention to $\Gamma_1^{(M=3)}$ which summarizes the contributions from the 
classes of Feynman diagrams involving three (truncated) external guons, as well as ghosts. 
Again, our first task is to compute the corresponding expression 
for the spin-factor. Eqs. (3.14) and (3.15) now give    
\begin{eqnarray}
Tr_L \Phi ^{\left[ 1 \right]} &=& D + 8\bar \xi _2 \xi _2 \bar \xi _3 \xi _3 
(\varepsilon ^2  \cdot p_3 \varepsilon ^3  \cdot p_2  - \varepsilon ^2  \cdot 
\varepsilon ^3 p_2  \cdot p_3 ) + 8\bar \xi _1 \xi _1 \bar \xi _3 \xi _3 (\varepsilon ^1  
\cdot p_3 \varepsilon ^3  \cdot p_1  - \varepsilon ^1  \cdot \varepsilon ^3 p_1  \cdot p_3 ) 
\nonumber\\&&
+ 8\bar \xi _1 \xi _1 \bar \xi _2 \xi _2 (\varepsilon ^1  \cdot p_2 \varepsilon ^2\cdot p_1  
- \varepsilon ^1  \cdot \varepsilon ^2 p_1  \cdot p_2 ) \nonumber\\&& 
 + \frac{{16}}{T}\bar \xi _1 \xi {}_1\bar \xi _2 \xi _2 \bar \xi _3 \xi _3
[(\varepsilon ^1\cdot \varepsilon ^2 \varepsilon ^3  \cdot p_1 - \varepsilon ^1\cdot\varepsilon
 ^3 \varepsilon ^2 \cdot p_1 )\delta (u_3 - u_2 )\nonumber\\&& 
+(\varepsilon ^1\cdot\varepsilon^3\varepsilon ^2  \cdot p_3- \varepsilon ^2\cdot\varepsilon ^3
\varepsilon ^1\cdot p_3 )\delta (u_2  - u_1 )  
 + T\varepsilon ^1  \cdot p_3 \varepsilon ^2  \cdot p_1 \varepsilon ^3  \cdot p_2 ].
\end{eqnarray}
  
The integration over the Grassmann variables can be systematically performed, yielding
the result
\begin{eqnarray}
&&F_1^{(M = 3)} (u_1 ,u_2 ,u_3 ;T) =  - \frac{{D - 2}}{T}\{\varepsilon ^1\cdot 
\varepsilon ^2 [\varepsilon ^3  \cdot p_1 \dot G(u_3 ,u_1 ) + \varepsilon ^3  \cdot p_2 
\dot G(u_3 ,u_2 )]\ddot G(u_1 ,u_2 ) +  \nonumber\\&& 
  + \varepsilon ^1  \cdot \varepsilon ^3 [\varepsilon ^2  \cdot p_1 \dot G(u_2 ,u_1 ) + 
\varepsilon ^2  \cdot p_3 \dot G(u_2 ,u_3 )]\ddot G(u_1 ,u_3 ) + \varepsilon ^2  \cdot 
\varepsilon ^3 [\varepsilon ^1  \cdot p_2 \dot G(u_1 ,u_2 ) + \varepsilon ^1  \cdot p_3 
\dot G(u_1 ,u_3 )] \nonumber\\&& 
  \times \ddot G(u_2 ,u_3 )\}  + \frac{{16}}{T}(\varepsilon ^1  \cdot \varepsilon ^2 
\varepsilon ^3  \cdot p_1  - \varepsilon ^1  \cdot \varepsilon ^3 \varepsilon ^2  
\cdot p_1 )\delta (u_3  - u_2 )\nonumber\\&& 
+ \frac{16}{T}(\varepsilon ^1  \cdot \varepsilon ^3 
\varepsilon ^2  \cdot p_3  - \varepsilon ^2  \cdot \varepsilon ^3 \varepsilon ^1  \cdot p_3 )
\delta (u_2  - u_1 ) + f. t., 
 \end{eqnarray}
where $f.t.$ stands for `terms with finite contribution'. Obviously the latter involve terms 
with $T$ to the 0th power or higher, equivalently, they involve more than one (external)
momentum variables. Let us reiterate that the finite terms should be computable through 
numerical methods that are currently being developed.
 
Substituting the above result in Eq. (3.10) we obtain 
\begin{eqnarray}
&&\Gamma _1^{(M = 3)} (p_1 ,p_2 ,p_3 ) =  - \frac{1}{2}(2\pi )^4 \delta ^{(4)} 
(p_1  + p_2  + p_3 )Tr_C (t_G^{a_3 } t_G^{a_2 } t_G^{a_1 } )\frac{{g^3 }}{{(4\pi )^{D/2} }}
\int\limits_0^\infty  {dTT^{1 - D/2} }\nonumber\\&& 
\int\limits_0^1 {du_3 \int\limits_0^{u_3 } {du_2 
\int\limits_0^{u_2 } {du_1}}}
\times \{ 4(D - 2)[\varepsilon ^1  \cdot \varepsilon ^2 \varepsilon ^3  \cdot p_2 
(u_2  - u_1 ) + \varepsilon ^1  \cdot \varepsilon ^3 \varepsilon ^2  \cdot p_1 (1 - 
(u_3  - u_1 ))] \nonumber\\&&
+ \varepsilon ^2  \cdot \varepsilon ^3 \varepsilon ^1  \cdot p_3 
(u_3  - u_2 )]
- 16(\varepsilon ^1  \cdot \varepsilon ^2 \varepsilon ^3  \cdot p_2  + \varepsilon ^1 
 \cdot \varepsilon ^3 \varepsilon ^2  \cdot p_1 )\delta (u_3  - u_2 )\nonumber\\&& 
- 16(\varepsilon ^1 .
\varepsilon ^3 \varepsilon ^2  \cdot p_1  + \varepsilon ^2  \cdot \varepsilon ^3 
\varepsilon ^1  \cdot p_3 )\delta (u_2- u_1 )+f.t.\}\nonumber\\&&
\times\exp \{  - \frac{{T\lambda ^2 }}{2}[(u_2  - u_1 )(1 - (u_2  - u_1 )) 
+ (u_3  - u_2 )(1 - (u_3  - u_2 ))\nonumber\\&& 
+ (u_3  - u_1 )(1 - (u_3  - u_1 ))]\}+permutations.
\end{eqnarray}
It is easy to see that the first term in the curly brackets takes care of the Feynman
diagrams involving three-gluon vertices while the other two, which carry the delta-functions,
collect the contributions from diagrams with one four-vertex. To further guide the
reader let us also mention that use was made of Eqs. (3.11) and (3.12). Accordingly,
the above expression refers to the specific ordering which enters these equations and 
underlies the particular integrations over the parameters $u_1$, $u_2$ and
$u_3$, an occurrence that will be rectified shortly. Finally, in the exponential 
factor we have set                                                                                                                                                                                                                                                                                                                    
\begin{equation}
p_1^2  = p_2^2  = p_3^2  = \lambda ^2,\quad
2p_1  \cdot p_2  = 2p_1 .p_3  = 2p_2  \cdot p_3  =  - \lambda ^2.
\end{equation}                     
Next, we make the variable change $u_2-u_1=x_2$ and $u_3-u_1=x_3$ which casts Eq. (4.8)
into the form
\begin{eqnarray}
&&\Gamma _1^{(M = 3)} (p_1 ,p_2 ,p_3 ) =- \frac{1}{2}(2\pi )^4 \delta ^{(4)}(p_1+ p_2+ p_3 )
Tr_C (t_G^{a_3 } t_G^{a_2 } t_G^{a_1 } )g\frac{{g^2 }}{{(4\pi )^2 }}\left( {4\pi 
\frac{{\mu ^2 }}{{\lambda ^2 }}} \right)^{2 - D/2}  \nonumber\\&&
\times \{  - 4(D - 2)a_D (\varepsilon ^1  \cdot \varepsilon ^2 \varepsilon ^3\cdot p_2
+ \varepsilon ^1  \cdot \varepsilon ^3 \varepsilon ^2  \cdot p_1  + \varepsilon ^2\cdot 
\varepsilon ^3 \varepsilon ^1  \cdot p_3 ) + 8b_D (\varepsilon ^1  \cdot \varepsilon ^2 
\varepsilon ^3  \cdot p_2\nonumber\\&&
+ 3\varepsilon ^1  \cdot \varepsilon ^3 \varepsilon ^2\cdot p_1+ 2\varepsilon ^2\cdot 
\varepsilon ^3 \varepsilon ^1  \cdot p_3 )\} \Gamma (2 - \frac{D}{2}) + f.t +permutations,
\end{eqnarray}
where we have introduced                                                                                                                                   
\begin{equation}
a_D  = 2^{2 - D/2} \int\limits_0^1 {dx_3 } \int\limits_0^{x_3 } {dx_2 } x_2 
\left[{x_2 (1 - x_2 ) + (x_3  - x_2 )(1 - x_3  + x_2 ) + x_3 (1 - x_3 )} \right]^{D/2 - 2} 
\end{equation}
and
\begin{equation}
b_D  = \int\limits_0^1 {dx_3 } \left( {x_3 (1 - x_3 )} \right)^{D/2 - 2}.
\end{equation}
Obviously $a_4={1\over^6}$ and $b_4=1$.

In order to obtain the final result we need to take into account contributions coming 
from all permutations of the variables $u_1,\,u_2,\,u_3$ and divide by 3! to compensate
for boson indistinguishability. The result can be easily obtained using the cyclic
properties of the trace. One, finally, obtains
\begin{eqnarray}                                                                                   
&&\Gamma _1^{(M = 3)} (p_1 ,p_2 ,p_3 ) = \frac{1}{2}(2\pi )^4 \delta ^{(4)} 
(p_1  + p_2  + p_3 )Tr_C (t_G^{a_3 } t_G^{a_2 } t_G^{a_1 } )g\frac{{g^2 }}{{(4\pi )^2 }}
\left( {4\pi \frac{{\mu ^2 }}{{\lambda ^2 }}} \right)^{2 - D/2}\nonumber\\&&\times 
\{ 4(D - 2)a_D (\varepsilon ^1  \cdot \varepsilon ^2 \varepsilon ^3  \cdot p_2 
 + \varepsilon ^1  \cdot \varepsilon ^3 \varepsilon ^2  \cdot p_1  + \varepsilon ^2  
\cdot \varepsilon ^3 \varepsilon ^1  \cdot p_3 ) - 16b_D (\varepsilon ^1  \cdot 
\varepsilon ^2 \varepsilon ^3  \cdot p_2 \nonumber\\&&
+ \varepsilon ^1  \cdot \varepsilon ^3 \varepsilon ^2  \cdot p_1  + \varepsilon ^2 
 \cdot \varepsilon ^3 \varepsilon ^1  \cdot p_3 )\} \Gamma (2 - \frac{D}{2}) + \, f.t. 
\end{eqnarray}
in agreement with the known result.           

\vspace{.5cm}
{\bf 4.3 The four gluon contribution (M=4) to the Effective Action}
\vspace{.3cm}

The computation in the present subsection pertains to a `world-line' diagram which
collectively accomodates all the classes of the contributing Feynman diagrams, i.e.
with zero, one  and two four-vertices (plus, of course, 
contributions from ghost diagrams). As our present analytic computations
refer to the divergent part, let us isolate the relevant contribution (terms with the
factor $1/T^2$) entering the expression for the spin-factor. We find
\begin{equation} 
Tr_L \Phi ^{\left[ 1 \right]}  = D + \frac{{32}}{{T^2 }}\bar \xi _4 \xi _4 \bar \xi _3 
\xi _3 \bar \xi _2 \xi _2 \bar \xi _1 \xi _1 (\varepsilon ^1  \cdot \varepsilon ^4 
\varepsilon ^2  \cdot \varepsilon ^3  - \varepsilon ^1  \cdot \varepsilon ^3 \varepsilon ^2  
\cdot \varepsilon ^4 )\delta (u_4  - u_3 )\delta (u_2  - u_1 )+ \, f.t.
\end{equation}
Integration over the Grassmann variables is a straightforward matter and gives                                                                            
\begin{eqnarray}
&&F^{(M = 4)} (u_4 ,u_3 ,u_2 ,u_1 ;T) = \frac{{D - 2}}{{T^2 }}[\varepsilon ^1  \cdot 
\varepsilon ^2 \varepsilon ^3  \cdot \varepsilon ^4 \ddot G(u_1 ,u_2 )\ddot G(u_3 ,u_4 )
\nonumber\\&& + \varepsilon ^1  \cdot \varepsilon ^3 \varepsilon ^2  \cdot \varepsilon ^4 
\ddot G(u_1 ,u_3 )\ddot G(u_2 ,u_4 ) + 
+ \varepsilon ^1  \cdot \varepsilon ^4 \varepsilon ^2  \cdot \varepsilon ^3 )
\ddot G(u_1 ,u_4 )\ddot G(u_2 ,u_3 )]\nonumber\\&& 
+ \frac{{32}}{{T^2 }}(\varepsilon ^1  \cdot 
\varepsilon ^4 \varepsilon ^2  \cdot \varepsilon ^3  - \varepsilon ^1  \cdot \varepsilon ^3 
\varepsilon ^2  \cdot \varepsilon ^4 )\delta (u_4  - u_3 )\delta (u_2  - u_1 ) + f.t.
\end{eqnarray}                                                                           
Substituting the above expression into Eq. (3.10) we get
\begin{eqnarray}
&&\Gamma _1^{(M = 4)} (p_1 ,p_2 ,p_3 ,p_4 ) =  - \frac{1}{2}(2\pi )^4 \delta ^{(4)} (p_1  + p_2
  + p_3  + p_4 )Tr_C (t_G^{a_4 } t_G^{a_3 } t_G^{a_2 } t_G^{a_1 } )\frac{{g_D ^4 }}
{{(2\pi )^{D/2} }} \nonumber\\&& 
  \times \int\limits_0^1 {du_4 } \int\limits_0^{u_4 } {du_3 } \int\limits_0^{u_3 } {du_2 
\int\limits_0^{u_2 } {du_1 } } \{ 4(D - 2)(\varepsilon ^1  \cdot \varepsilon ^2 \varepsilon ^3 
 \cdot \varepsilon ^4  + \varepsilon ^1  \cdot \varepsilon ^3 \varepsilon ^2  \cdot \varepsilon
 ^4+\varepsilon ^1\cdot \varepsilon ^4 \varepsilon ^2\cdot \varepsilon ^3 )\nonumber\\&& 
  - 4(D - 2)\varepsilon ^1  \cdot \varepsilon ^2 \varepsilon ^3  \cdot \varepsilon ^4 
[\delta (u_2  - u_1 ) + \delta (u_4  - u_3 )] - 4(D - 2)\varepsilon ^1  \cdot \varepsilon ^4 
\varepsilon ^2  \cdot \varepsilon ^3 \delta (u_3  - u_2 ) +  \nonumber\\&& 
  + 4(D - 2)\varepsilon ^1  \cdot \varepsilon ^2 \varepsilon ^3  \cdot \varepsilon ^4 
\delta (u_2  - u_1 )\delta (u_4  - u_3 ) + 32(\varepsilon ^1  \cdot \varepsilon ^4 
\varepsilon ^2  \cdot \varepsilon ^3  - \varepsilon ^1  \cdot \varepsilon ^3 
\varepsilon ^2  \cdot \varepsilon ^4 )\nonumber\\&&\times
\delta (u_2  - u_1 )\delta (u_4  - u_3 )\}  
\left[ {\sum\limits_{n = 1}^4 {\sum\limits_{m = n + 1}^4 {p_n  
\cdot p_m G(u_n ,u_m )} } } \right]^{D/2 - 2} \Gamma (2 - \frac{D}{2})\nonumber\\&& 
+f.t.+permutations. 
\end{eqnarray}
One can easily verify that the first term inside the curly brackets represents contributions 
corresponding to the Feynman diagrams with no four-vertices, the next two to those with one
and the last to those with two. Of course, the above expression pertains to a particular
ordering of the variables $u_1,\,u_2,\,u_3,\,u_4$ as reflected in the explicit delta functions 
which make their entrance.

Performing the parametric integrations, in the specific ordering that appears in Eq. (4.16), 
one obtains
\begin{eqnarray}
&& \Gamma _1^{(M = 4)} (p_1 ,p_2 ,p_3 ,p_4 ) =  - \frac{1}{2}(2\pi )^4 \delta ^{(4)} 
(p_1  + p_2  + p_3  + p_4 )Tr_C (t_G^{a_4 } t_G^{a_3 } t_G^{a_2 } t_G^{a_1 } )
\frac{{g_D ^4 }}{{(2\pi )^{D/2} }} \nonumber\\&& 
  \times \{ 4(D - 2)A_D (\varepsilon ^1  \cdot \varepsilon ^2 \varepsilon ^3  \cdot 
\varepsilon ^4  + \varepsilon ^1  \cdot \varepsilon ^3 \varepsilon ^2  \cdot \varepsilon ^4  
+ \varepsilon ^1  \cdot \varepsilon ^4 \varepsilon ^2  \cdot \varepsilon ^3 ) - 4(D - 2)
(B_D- C_D )\varepsilon ^1\cdot \varepsilon ^2\varepsilon ^3\cdot\varepsilon ^4-\nonumber\\&& 
  - 4(D - 2)D_D \varepsilon ^1  \cdot \varepsilon ^4 \varepsilon ^2  \cdot \varepsilon ^3 
 + 32C_D (\varepsilon ^1  \cdot \varepsilon ^4 \varepsilon ^2  \cdot \varepsilon ^3  - 
\varepsilon ^1  \cdot \varepsilon ^3 \varepsilon ^2  \cdot \varepsilon ^4 )\} \Gamma (2 - 
\frac{D}{2}) +f.t.\nonumber\\&&
\quad +permutations, 
 \end{eqnarray}
where we have set
\begin{eqnarray}
&&A_D\equiv\int\limits_0^1 {du_4 } \int\limits_0^{u_4 } {du_3 } \int\limits_0^{u_3 } {du_2 }
 \int\limits_0^{u_2 } {du_1 } \left[ {\sum\limits_{n = 1}^4 {\sum\limits_{m = n + 1}^4 {p_n 
 \cdot p_m G(u_n ,u_m )} } } \right]^{D/2 - 2} ,\nonumber\\&&
B_D\equiv\int\limits_0^1 {du_4 } \int\limits_0^{u_4 } {du_3 } \int\limits_0^{u_3 } {du_2 }
 \int\limits_0^{u_2 } {du_1 } [\delta (u_2  - u_1 ) + \delta (u_4  - u_3 )]\left[ 
{\sum\limits_{n = 1}^4 {\sum\limits_{m = n + 1}^4 {p_n  \cdot p_m G(u_n ,u_m )} } } 
\right]^{D/2 - 2} ,\nonumber\\&&
C_D\equiv\int\limits_0^1 {du_4 } \int\limits_0^{u_4 } {du_3 } \int\limits_0^{u_3 } {du_2 } 
\int\limits_0^{u_2 } {du_1 } \delta (u_2  - u_1 )\delta (u_4  - u_3 )\left[ 
{\sum\limits_{n = 1}^4 {\sum\limits_{m = n + 1}^4 {p_n  \cdot p_m G(u_n ,u_m )} } } 
\right]^{D/2 - 2} ,\nonumber\\&&
D_D\equiv\int\limits_0^1 {du_4 } \int\limits_0^{u_4 } {du_3 } \int\limits_0^{u_3 } 
{du_2 } \int\limits_0^{u_2 } {du_1 } \delta (u_3  - u_2^{} )\left[ {\sum\limits_{n = 1}^4 
{\sum\limits_{m = n + 1}^4 {p_n  \cdot p_m G(u_n ,u_m )} } } \right]^{D/2 - 2}.
\end{eqnarray}
One trivially finds $A_4= \frac{1}{6}$, $B_4= \frac{3}{4}$, $C_4  = \frac{1}{2}$ 
and $D_4  = \frac{1}{4}$.

The remaing step is to perform all reorderings of the $u$ variables and divide by 4!. In this
way one arrives at the final expression
\begin{eqnarray}
&&\Gamma _1^{(M = 4)}(p_1 ,p_2 ,p_3 ,p_4 )=- \frac{1}{2}(2\pi )^4 \delta ^{(4)} (p_1+ p_2
  + p_3  + p_4 )Tr_C (t_G^{a_4 } t_G^{a_3 } t_G^{a_2 } t_G^{a_1 } )\frac{{g_D ^4 }}
{{(2\pi )^{D/2} }} \nonumber\\&& 
  \times \{ 4(D - 2)A_D (\varepsilon ^1  \cdot \varepsilon ^2 \varepsilon ^3  \cdot 
\varepsilon ^4  + \varepsilon ^1  \cdot \varepsilon ^3 \varepsilon ^2  \cdot \varepsilon ^4  
+\varepsilon ^1\cdot \varepsilon ^4\varepsilon ^2\cdot \varepsilon ^3 )-2(D-2)\nonumber\\&& 
\times(B_D  - C_D  + D_D )(\varepsilon ^1  \cdot \varepsilon ^2 \varepsilon ^3  
\cdot \varepsilon ^4  + \varepsilon ^1  \cdot \varepsilon ^4 \varepsilon ^2  \cdot 
\varepsilon ^3 ) + 16C_D (\varepsilon ^1  \cdot \varepsilon ^2 \varepsilon ^3  \cdot 
\varepsilon ^4  + \varepsilon ^1  \cdot \varepsilon ^4 \varepsilon ^2  \cdot \varepsilon ^3 ) 
\nonumber\\&& 
  - 32C_D \varepsilon ^1  \cdot \varepsilon ^3 \varepsilon ^2  \cdot \varepsilon ^4 \} 
\Gamma (2 - \frac{D}{2}) + f.t., 
 \end{eqnarray}
which is in full agreement with the known results.

\section{Concluding comments}

Given the schemes pioneered by Bern and Kosower [6] and reformulated by Strassler [13] based on
string and world-line agents, respectively, and which aim at expediting perturbative 
computations in QCD both economically and efficiently, it becomes important to assess the 
relevant merits of yet another competitive proposal advanced in the present paper
which utilizes the Polyakov world-line path integral. Directing, to begin with, our comments 
towards making comparisons with Strassler's approach we could say that
the basic difference between the two world-line based schemes is how the disentanglement,
between the weight factor pertaining to the spin of the propagating particle-like object on a 
given path and the dynamical factor represented by the Wilson line (loop), is accomplished. In 
Strassler's case this task is confronted by using super-particle degrees of
freedom (one dimensional) and generates a term in the corresponding Lagrangian of the form
$\psi^\mu F_{\mu\nu}\psi^\nu$.
In the Polyakov (world-line) version, on the other hand, the issue is addressed via 
the introduction of the spin-factor. We believe that 
the separation, featured by the latter scheme, between `geometrical' characteristics of paths 
on the one hand and dynamics -as embodied in the Wilson line(loop) factor- on the other,
leads to an organization of the path integral expression which further facilitates 
the `efficiency factor' for performing perturbative computations. In particular, it offers
a unified base for treating, spinors, gauge fields and ghosts; all one has to do is 
adjust the master formula, which yields the computational rules, to the appropriate spin factor.
Moreover, it lends itself to  straightforward extensions for applications to processes 
involving {\it open} fermionic world-lines, as established in Refs [30, 31]. Referring, 
finally, to the string-based approach of Bern and Kosower we note that 
the pinching issue, which arises in the non-abelian case and whose 
confrontation requires the application of a certain set of mnemonic rules, does not arise in
our approach. We expect to further demostrate the virtues of the Polyakov world-line 
path integral scheme toward the calculation of two gluon loop contributions 
to the effective action by generating the corresponding master formulas.      
\vspace*{1cm}\\
\textbf{\large{Acknowledgements}}\\
One of us (S. D. A) acknowledges financial support from the Greek State Scholarships 
Foundation (I.K.Y.). A. I. K. and C. N. K. acknowledge the support from the General 
Secretariat of Research and Technology of the University of Athens.
\newpage

\appendix
\setcounter{section}{0}
\addtocounter{section}{1}
\section*{Appendix}
\setcounter{equation}{0}
\renewcommand{\theequation}{\thesection.\arabic{equation}}
In this Appendix we shall pay closer attention to the spin factor with respect to both
carrying out the path integral in Eq. (3.3) and establishing the result encoded in
Eqs. (3.14) and (3.15). Looking at the identity given by Eq. (2.11) we present the
proper (regularized) expression for the tensor $\omega _{\mu \nu }$ reads as follows
\begin{equation}
\int\limits_0^T {dt\omega _{\mu \nu } } [\dot x(t)] = \mathop {\lim }\limits_{\varepsilon  
\to 0} \frac{1}{4}\int\limits_{ - \varepsilon }^\varepsilon  {ds\int\limits_0^T {dt_2 } } 
\int\limits_0^T {dt_1 } [\ddot x_\mu  (t_2 )\dot x_\nu  (t_1 ) - \ddot x_\nu  (t_2 )
\dot x_\mu  (t_1 )]\delta (t_2  - t_1  - s).
\end{equation}
Now, if the functions (on the line) $x_\mu (t)$ are
infinitely differentiable, then we can, once taking into account that $\left|{t_2-t_1}\right|
<\varepsilon$, write $\ddot x_\mu(t_2) = \ddot x_\mu(t_1)+{\cal O}(s)$ as well as
$\dot x_\nu(t_1)=\dot x_\nu(t_2)+{\cal O}(s)$ and immediately conclude that
\begin{equation}
\int\limits_0^T {dt\omega _{\mu \nu } [\dot x(t)] = \frac{T}{2}} \int\limits_0^T {dt[\ddot 
x_\mu}(t)\dot x_\nu(t) -\ddot x_\nu(t)\dot x_\mu(t)].
\end{equation}
Otherwise, one should use the limiting expression according to (A.1) when performing 
manipulations that involve the spin factor.

Let us proceed with the computation of the path integral entering Eq. (3.3). We set
\begin{equation}                  
{\rm I}_{\mu \nu}=\int{d^4 a\int\limits_{x(0)= x(T)= a}{Dx(t)}}\Phi^{[1]}[\dot x(t)]_{\mu\nu} 
\exp\{- S[x]\}, 
\end{equation}
where
\begin{equation}
S[x] = \frac{1}{4}\int\limits_0^T {dt\dot x^2 }(t) -i\sum\limits_{n = 1}^M {\hat k(t_n }) 
\cdot x(t_n ).
\end{equation}                                          

To compute ${\rm I}_{\mu \nu}$ we make the variable change $x\to x+ x^{cl}$, where $x^{cl}$
is a solution of the classical equation of motion resulting from the above action.
Specifically, we have
\begin{equation}
\ddot x_\mu^{cl}(t)=- 2i\sum\limits_{n = 1}^M {\hat k(t_n } )\delta (t-t_n)\Rightarrow 
x_\mu^{cl}(t)= 2i\sum\limits_{n =1}^M {\hat k(t_n})\Delta(t,t_n )+ a,
\end{equation}
where we have employed the Green's function 
\begin{equation}                                                 
\Delta(t,t')=\frac{{t(T-t')}}{T}\theta(t'-t)+\frac{{t'(T-t)}}{T}\theta(t-t'),\quad
\Delta(0,t')=\Delta(T,t')=0.
\end{equation}

The new action functional is now specified by 
\begin{equation}
S[x] \to \frac{1}{4}\int\limits_0^T {dt\dot x^2 } (t) + S[x^{cl} ],
\end{equation}
where                                
\begin{equation}
S[x^{cl} ]= \sum\limits_{n = 1}^M {\sum\limits_{m = 1}^M {\hat k(t_n } } ) 
\cdot \hat k(t_m )\Delta (t_n ,t_m ) - i\sum\limits_{n = 1}^M {p_n }  \cdot a.
\end{equation}
We immediately observe that
integration over $\alpha$ (translational zero modes) leads to momentum conservation which
enters Eq. (3.7). The rest of the expression for $S[x^{cl}]$ produces the terms
entering the exponential factor in the same equation.

Turning our attention to the spin factor we first note that the variable change           
$x \to x + x^{cl}$ leads to
\begin{equation}
\int\limits_0^T{dt}\omega_{\mu \nu}[\dot x]\to \int\limits_0^T{dt\omega_{\mu \nu}}[\dot x^{cl}]
+\frac{T}{2}\int\limits_0^T {dt[\ddot x_\mu}(t)\dot x_\nu(t) - \ddot x_\nu(t)\dot x_\mu(t)],
\end{equation}
having taken into account that the contours $x(t)$ are to be integrated with respect to
a quadradtic action functional (cf. Eq. (A.7)), which implies, cf. Eqs. (A.1) and (A.5), that 
mixed terms in $x$ and $x^{cl}$ drop out. Let us finally note that
for paths that are infinitely differentiable, in which case Eq. (A.2) strictly holds true,
the integration of the spin factor with respect to the quadratic action functional yields 
unity. All this leads to the following result as far as performing the path integral
in Eq. (A.3) is concerned.
\begin{eqnarray}
&& I_{\mu \nu }  = (2\pi )^D \delta ^{(D)} \left(\sum\limits_{n = 1}^M {p_n }\right)
\frac{1}{{(4\pi T)^{D/2} }}\Phi ^{[1]} [\dot x^{cl} ]_{\mu \nu }\nonumber\\&& 
\times \exp \left[{\sum\limits_{n < m} {p_n  \cdot p_m G(t_n ,t_m ) + \sum
\limits_{n \ne m} {\bar \xi _n \xi _n \varepsilon ^n  \cdot p_m \partial _n G(t_n ,t_m)}}}
\right.\nonumber\\&&\quad +\left. \frac{1}{2}
\sum\limits_{n \ne m} {\bar \xi _n \xi _n \bar \xi _m \xi _m \varepsilon ^n  \cdot 
\varepsilon ^m \partial _n \partial _m G(t_n ,t_m )} \right].  
\end{eqnarray}
The above result explicitly demonstrates our assertion that the overall contribution from 
the spin factor is exclusively determined by those points on a given path where a momentum
is imparted via the action of an external gauge field.
                                         
The final result is obtained once we substitute (A.5) into (A.1). We get
\begin{eqnarray}
\int\limits_0^T {dt\omega _{\mu \nu } } [\dot x^{cl} ]&=& - 2\sum\limits_{n = 1}^M 
{\bar \xi _n } \xi _n (\varepsilon _\mu ^n p_{n,\nu }  - \varepsilon _\nu ^n p_{n,\mu } ) 
+ \sum\limits_{n = 0}^M {\sum\limits_{m = 0}^M {\bar \xi _n \xi _n \bar \xi _m \xi _m } } 
(\varepsilon_\mu^n\varepsilon_\nu^m-\varepsilon_\nu^n\varepsilon_\mu ^m)\nonumber\\&&\quad
\times\int\limits_{ - \varepsilon }^\varepsilon  {ds\frac{\partial }{{\partial t_n }}} 
\delta(t_n-t_m-s).
\end{eqnarray}                                                  
The correct handling of the last term follows once we take into consideration that,
first, $m\ne n$ on account of the Grassmann variables and, second, it is
to be integrated in consistence with the time ordering implicit in Eq. (3.1) of
the text. Specifically, we have
\begin{eqnarray}
&&...\int\limits_0^T {dt_{n + 1} } \int\limits_0^T {dt_n } \int\limits_0^T {dt_{n - 1} } 
\theta (t_{n + 1}  - t_n )\theta (t_n  - t_{n - 1} )\int\limits_{ - \varepsilon }^\varepsilon  
{ds\frac{\partial }{{\partial t_n }}} \delta (t_n  - t_m  - s)...=\nonumber\\ &&
= ...\int\limits_0^T {dt_{n + 1} } \int\limits_0^T {dt_n } \int\limits_0^T {dt_{n - 1} } 
\theta (t_{n + 1}-t_n)\theta (t_n-t_{n -1})\nonumber\\&&\quad \times
[2\delta _{n + 1,m} \delta (t_{n + 1}-t_n)- 2\delta_{m,n-1}\delta (t_n-t_{n-1})]...,
\end{eqnarray}
which allows us to return to Eq. (A.11) and infer that
\begin{eqnarray}
 \int\limits_0^T {dt\omega_{\mu \nu } [\dot x^{cl}]}&=&- \frac{i}{2}} \int\limits_0^T 
{dt(J \cdot \omega ^{cl} )_{\mu \nu }= - 2\sum\limits_{n = 0}^M {\bar \xi _n \xi _n} 
(\varepsilon _\mu ^n p_{n,\nu }- \varepsilon _\nu ^n p_{n,\mu } )- \nonumber\\&& \quad\quad
 - 4\sum\limits_{n = 1}^M 
{\bar \xi _{n + 1} \xi _{n + 1} \bar \xi _n \xi _n } (\varepsilon _\mu ^{n + 1} 
\varepsilon _\nu ^n-\varepsilon_\nu ^{n + 1}\varepsilon _\mu ^n )\delta(t_{n + 1}-t_n )
\end{eqnarray}
With the above results in place, Eqs. (3.14) and (3.15) in the text follow directly.

The careful course of reasoning we have followed in this appendix can be circumvented by 
the more formal line of procedure adopted in the text. Thus, the validity of the
aformentioned equations can be established once we observe that  
\begin{eqnarray}
&&\exp \left[{-i\sum\limits_{n=1}^M {\hat k(t_n ) \cdot x(t_n )} } \right]\int\limits_0^T 
{dt\frac{\delta }{{\delta s_{\mu \nu }(t)}}} \exp \left[ {i\sum\limits_{n = 1}^M {\hat k(t_n ) 
\cdot x(t_n )} } \right] \nonumber\\&& 
  =  - \mathop {\lim }\limits_{\varepsilon  \to 0} \int\limits_0^T {dt\sum\limits_{n = 1}^M 
{\sum\limits_{m = 1}^M {\int\limits_{ - \varepsilon }^\varepsilon  {dss\hat k_\mu  (t_n }}}}
 )\delta (t_n  - t - \frac{s}{2})\hat k_\nu  (t_m )\delta (t_m-t+\frac{s}{2}) \nonumber\\&& 
  = \frac{1}{4}\mathop {\lim }\limits_{\varepsilon  \to 0} \int\limits_0^T {dt} 
\int\limits_{ - \varepsilon }^\varepsilon  {dss\ddot x_\mu ^{cl} } (t + \frac{s}{2})
\ddot x_\nu ^{cl} (t - \frac{s}{2})\nonumber\\&& 
= \mathop {\lim }\limits_{\varepsilon  \to 0} 
\frac{1}{8}\int\limits_{ - \varepsilon }^\varepsilon  {ds\int\limits_0^T {dt_2 } } 
\int\limits_0^T {dt_1 } [\ddot x_\mu  (t_2 )\dot x_\nu  (t_1 ) - \ddot x_\nu  (t_2 )
\dot x_\mu  (t_1 )]\delta (t_2  - t_1  - s). 
\end{eqnarray}








\end{document}